\documentclass[3p,times]{elsarticle}

\usepackage{ecrc}

\volume{00}

\firstpage{1}

\journalname{Nuclear Physics A}

\runauth{}

\jid{nupha}

\usepackage{amssymb}
\usepackage[figuresright]{rotating}

\newcommand\be{\begin{equation}}
\newcommand\ee{\end{equation}}
\newcommand{\bea}{\begin{eqnarray}}
\newcommand{\eea}{\end{eqnarray}}
\newcommand{\gton}{\mathrel{\lower.9ex \hbox{$\stackrel{\displaystyle 
>}{\sim}$}}} 
\newcommand{\lton}{\mathrel{\lower.9ex \hbox{$\stackrel{\displaystyle 
<}{\sim}$}}}

\newcommand{\vx}{{\vec x}}

\begin{document}

\begin{frontmatter}



\dochead{}

\title{Realistic medium-averaging in radiative energy loss}


\author{Denes Molnar}
\author{Deke Sun}

\address{Physics Department, Purdue University, West Lafayette, IN 47907}

\begin{abstract}
We present results from a jet energy loss 
calculation 
using the Gyulassy-Levai-Vitev (GLV) formalism
and bulk medium evolution from the covariant transport MPC. 
At both RHIC and LHC energies we find that
realistic transverse expansion strongly reduces 
elliptic flow at high $p_T$ compared to calculations with transversely
``frozen'' profiles. We argue that this is a generic feature of GLV energy 
loss.
Transverse expansion also leads to stronger high-$p_T$ suppression, 
while fluctuations in energy loss with the location of scattering
centers weaken the suppression. But, unlike the reduction of $v_2$,
these effects
nearly disappear once $\alpha_s$ is 
adjusted to reproduce $R_{AA}$ in central collisions.
\end{abstract}


\end{frontmatter}

% \linenumbers

\section{Introduction }
\label{Sc:Intro}
Understanding parton energy loss in ultrarelativistic heavy-ion reactions has 
been the focus of considerable recent theoretical effort. A variety of
phenomenological approaches
(e.g., \cite{Jia_Eloss,Liao_Shuryak_Eloss,Betz_Eloss}) 
formulate the problem in terms of 
a local energy loss rate $dE/dL = - f(E(L),T(L), L)$
along the 
Eikonal parton
trajectory, given by the
local temperature, position, and parton energy.
In the small-coupling regime, more rigorous treatment is possible based on
perturbative QCD\cite{BDMPS,WS,GLV}.
This includes quantum interference effects and also 
fluctuations, namely, energy loss along a given jet trajectory becomes a 
stochastic variable that is in general a 
function of the scattering and emission history of the jet.

A critical step in computing heavy-ion observables from any energy loss model 
is spatial and temporal averaging over the bulk medium formed in the 
collision. We employ here the Gyulassy-Levai-Vitev (GLV) 
framework\cite{GLV} 
in which a high-energy parton loses energy
through gluon radiation induced by interactions with static Yukawa
scatterers in the medium. It is 
natural to combine this approach with parton transport for the bulk evolution,
such as Molnar's Parton Cascade\cite{MPC} (MPC), because it directly
provides scattering center information.

Our approach is similar to recent work by
Buzzatti and Gyulassy\cite{CUJET1p0}, 
but with a few key differences. Unlike \cite{CUJET1p0}, 
at present we only focus on light partons, and do not include multiple 
gluon radiation, elastic energy loss, or 
energy loss fluctuations due to variations in radiated gluon 
momentum.  
However, 
we {\em do} include realistic 3D medium evolution with both longitudinal and 
transverse
expansion, which turns out to influence energy loss and, especially, 
elliptic flow. 

\section{Radiative energy loss and medium averaging}

We consider here the leading $n=1$ (single scattering) term in the GLV 
opacity expansion of the radiated gluon spectrum\cite{GLV}
\be
x\frac{dN^{(1)}}{dx\, d^2 {\bf k}} =
\frac{C_R \alpha_s}{\pi^2} \chi 
\int d^2 {\bf q} \, 
\frac{\mu^2(z)}{\pi ({\bf q}^2 + \mu^2(z))^2}
\, \frac{{\bf k}{\bf q}}{{\bf k}^2 ({\bf k} - {\bf q})^2}
\left[ 1 -   \cos \left (\, \frac{({\bf k} - {\bf q})^2 z}{2xE} \right)\right]
\label{GLV1}
\ee
where the original hard scattering is at $z=0$, $\mu(z)$ is the local
Debye screening mass, $\sigma = 9\pi \alpha_s^2/(2\mu^2)$ is the 
(screened) total $gg\to gg$ scattering cross section, 
and $\chi = \int dz\, \rho, \sigma$ is the opacity.
We integrated this spectrum numerically with kinematic bounds
$k < x E$, $q < \sqrt{6ET}$, and $x E \ge \mu$ to obtain a momentum-averaged 
energy loss 
$\Delta E^{(1)}(z) = \int dx\, d^2{\bf k}\,
E x\, (dN^{(1)}/dx\, d^2 {\bf k})$ for fixed $z$,
i.e., retained energy loss fluctuations due to variations in $z$ only.
The probability for the scattering to occur at $z$ is
$p(z) = \rho(z) \sigma(z) / \chi$, so the fully averaged energy loss is
$\Delta E^{(1)} = \int dz \, p(z)\, \Delta E^{(1)}(z)$.
\begin{figure}[t]
\leavevmode
\begin{center}
\includegraphics[height=50mm]{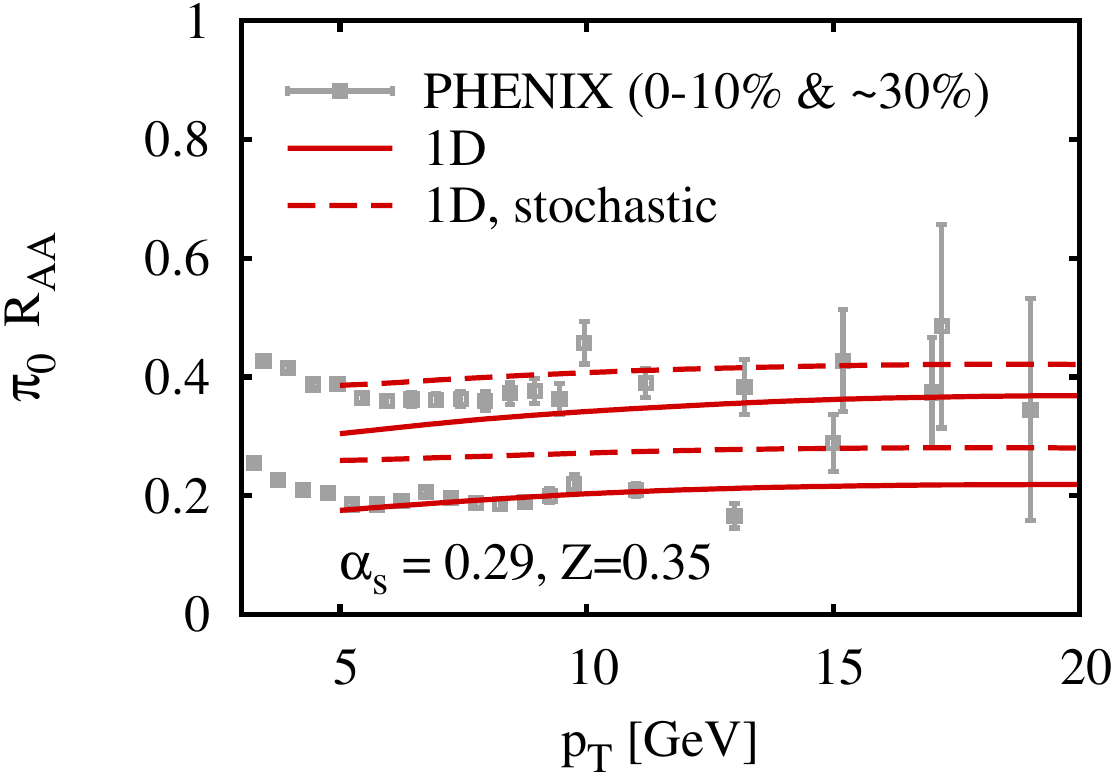}
\hskip 0cm
\includegraphics[height=50mm]{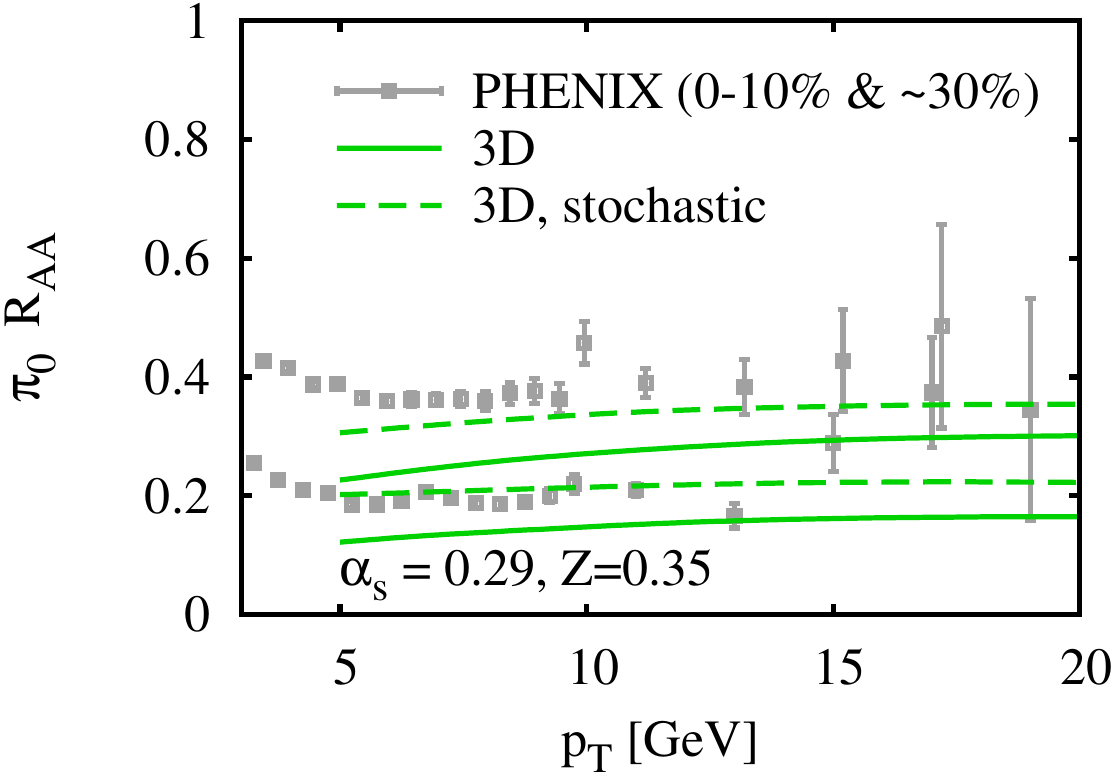}
\caption{Neutral pion suppression factor $R_{AA}$ at midrapidity 
in central and mid-central 
Au+Au at $\sqrt{s_{NN}} = 200$ GeV from calculations with GLV energy loss.
{\em Left:} results for a 
medium undergoing longitudinal Bjorken expansion only.
{\em Right:} results for realistic medium evolution with both longitudinal and
transverse expansion (see text for details). 
In both cases, two scenarios are considered,
one based on the average energy loss along the jet pass (solid), and one
utilizing stochastic energy loss $\Delta E(z)$ (dashed).
Data from PHENIX (boxes) are also 
shown\cite{PHENIX_pi0_y2008} to guide the eye.}
\label{Fig:R_AA}
\end{center}
\end{figure}
\begin{figure}[ht]
\leavevmode
\begin{center}
\includegraphics[height=50mm]{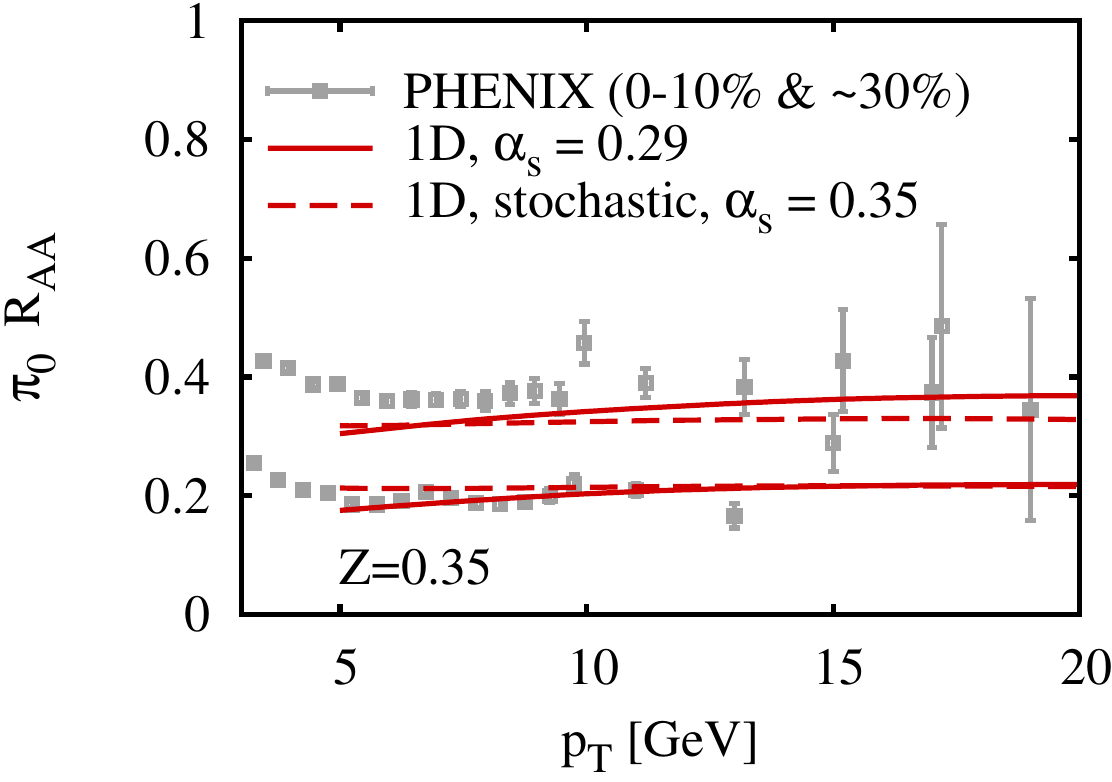}
\hskip 0cm
\includegraphics[height=50mm]{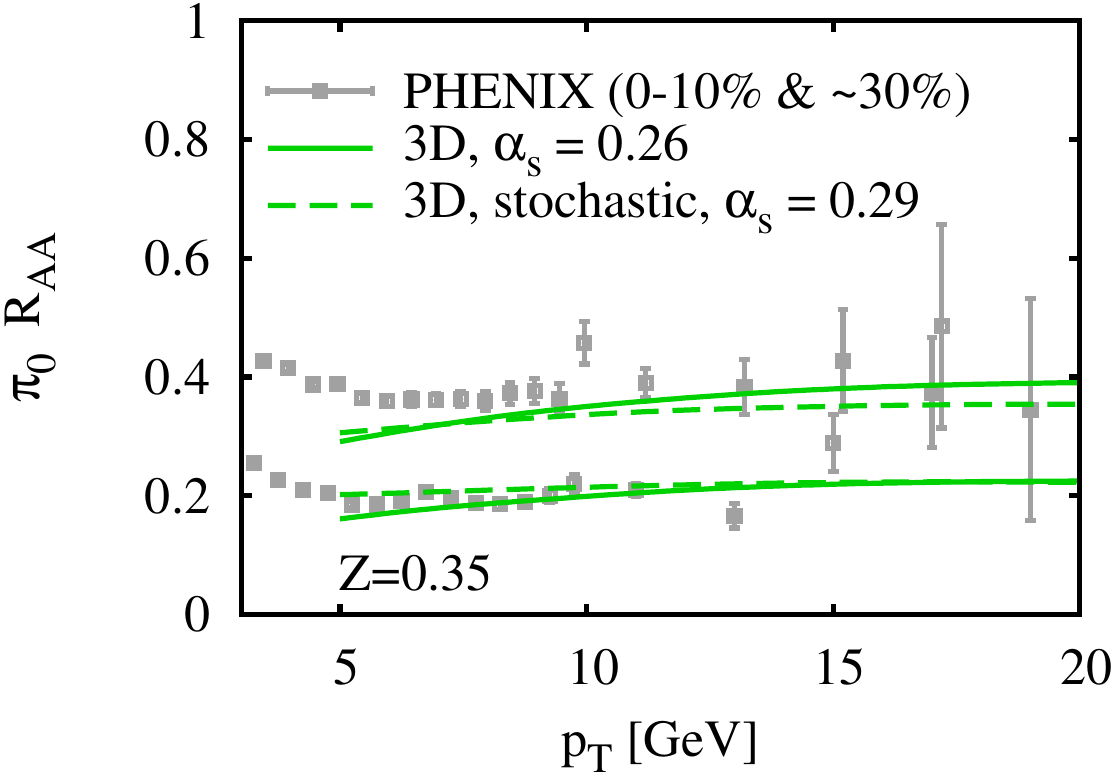}
\caption{Same as Fig.~\ref{Fig:R_AA} but with a slight tuning of 
$\alpha_s$ in each of the four scenarios to match the data above $p_T \gton 8$
 GeV in {\em central} collisions. After the tuning, 
all scenarios give practically
the same result.}
\label{Fig:R_AA_tuned}
\end{center}
\end{figure}

As customary, in non-static media we reinterpret $\rho(z)$
in the GLV formula as $\rho(z,t=t_0 + z)$ along the parton trajectory.
The density evolution was obtained from the parton
transport MPC, 
employing $2\to 2$ interactions for massless gluons. The scattering rate
was adjusted to generate substantial $v_2(p_T\approx 3\ {\rm GeV}) \sim 0.25$ 
in collisions with $b=8$ fm impact parameter,
and we set growing $\sigma_{gg\to gg} \propto \tau^{2/3}$ to keep
the shear viscosity to entropy ratio approximately constant\cite{minvisc}.
Initial conditions
for Au+Au at $\sqrt{s_{NN}} = 200$~GeV and 
Pb+Pb at $\sqrt{s_{NN}} = 2.76$~TeV were based on 
diffuse Woods-Saxon nuclei. For the transverse density, 
binary collision profiles were used, while the impact parameter dependence of 
rapidity densities $dN(b)/dy$ was proportional to $N_{part}$
with $dN(0)/dy = 1100$ (Au+Au) and 2400 (Pb+Pb) to match the
observed charged particle yields. Because we 
are only interested in observables
at midrapidity, we set up boost invariant conditions 
in the coordinate rapidity window $|\eta| < 5$, with formation time
$\tau_0 = 0.6$~fm.

Using tabulated densities from the transport, we set the local
temperature assuming a massless gas of gluons $\rho \approx 2 T^3$ and
the Debye mass via $\mu = g T \approx 2T$. At early times 
$\tau < \tau_0$ we assume linear density build-up\cite{CUJET1p0}
$\rho = \tau \rho(\vx_T, \tau_0) / \tau_0$.
We roughly account for additional energy loss off
dynamical (recoiling) scattering centers\cite{DGLV_dyn} 
[$(q^2+\mu^2)^2 \to q^2(q^2+\mu^2)$ in (\ref{GLV1})]
and
elastic scattering\cite{GLV_rad+el} 
via rescaling our opacities $\chi \to \chi/Z$ with $Z = 0.35$.
Initial jet momentum distributions in p+p, Au+Au and Pb+Pb
were computed from leading-order (LO) perturbative QCD with one-loop running
coupling $\alpha_s(Q^2)$, using CTEQ5L parton distribution function 
parameterizations with $Q^2 = p_{T,parton}^2$.
Nuclear effects such as shadowing were ignored but isospin
(proton-neutron difference) was included.
After energy loss, jets were fragmented
independently
using LO BKK95 fragmentation function parameterizations with scale factor
$Q^2 = p_{T,hadron}^2$ and we assumed
$\pi_0 = (\pi^+ + \pi^-) / 2$ for the neutral pion yield.
This procedure
reproduces high-$p_T$ $\pi_0$ and charged particle 
spectra in p+p at RHIC and LHC with modest 
$K_{NLO} \approx 2.5$ to account for higher-order contributions.

Below we 
focus on two basic high-$p_T$ observables for neutral pions at 
midrapidity, 
the nuclear suppression factor 
$R_{AA}$ and the momentum anisotropy (elliptic flow)
$v_2 = \langle \cos 2\phi \rangle_{p_T}$.
Only energy loss was considered, i.e.,
contributions by the radiated gluons to the final spectrum
and feedback on the bulk medium due to the jet were ignored.

%%%
\section{Main results}

\begin{figure}[t]
\leavevmode
\begin{center}
\includegraphics[height=50mm]{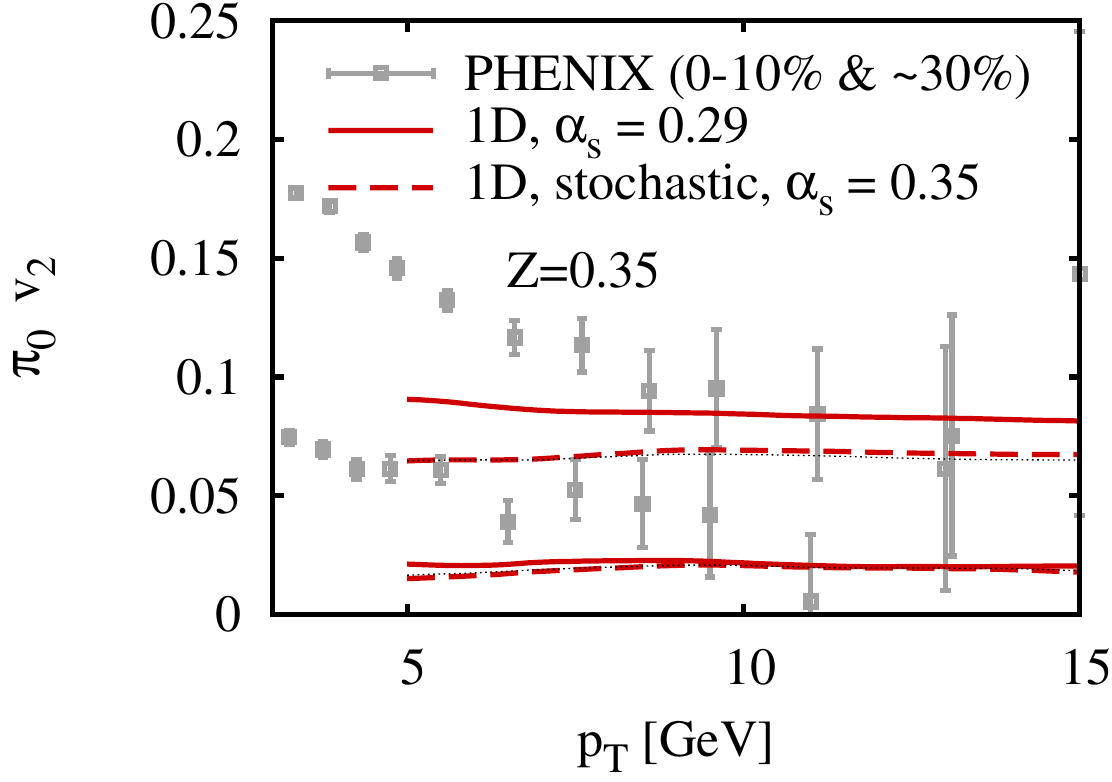}
\hskip 0cm
\includegraphics[height=50mm]{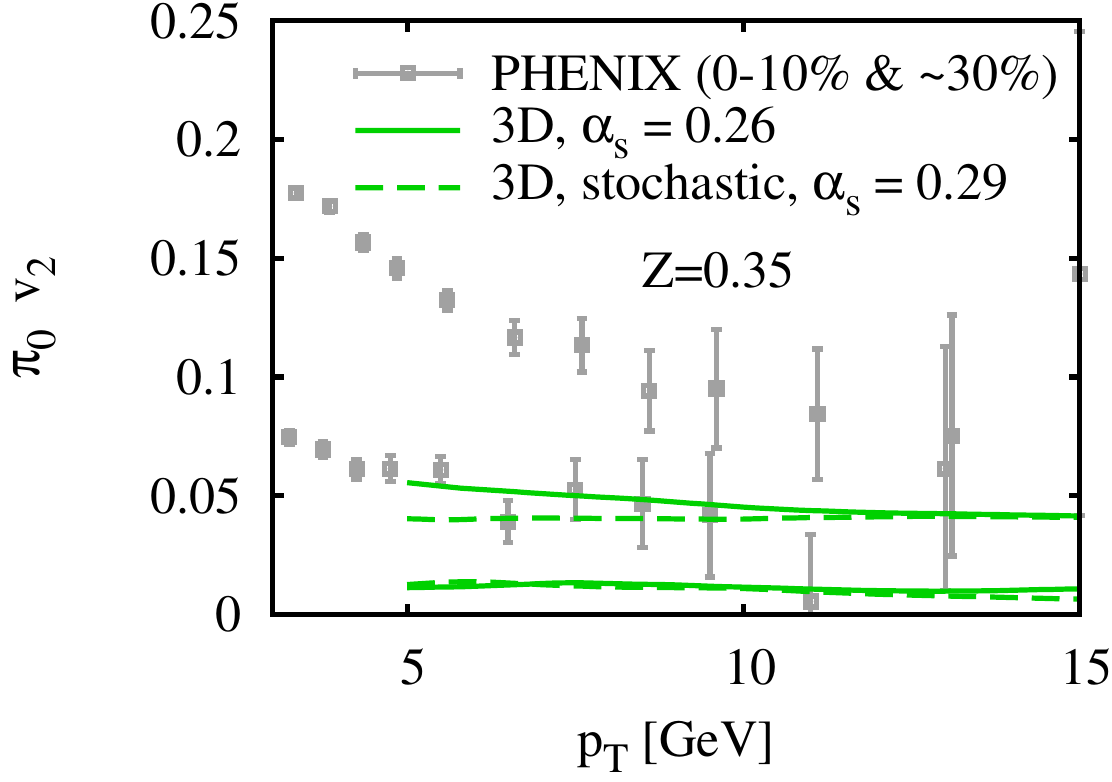}
\caption{Differential $\pi_0$ elliptic flow $v_2(p_T)$ at midrapidity 
in central and mid-central 
Au+Au at $\sqrt{s_{NN}} = 200$ GeV from calculations with GLV energy loss.
The value of $\alpha_s$ was tuned to reproduce 
$R_{AA}$ in central collisions (cf. Fig.~\ref{Fig:R_AA_tuned}).
{\em Left:} results for a 
medium undergoing longitudinal Bjorken expansion only.
{\em Right:} results for realistic medium evolution with both longitidunal and
transverse expansion (see text for details). 
In both cases, two scenarios are considered,
one is based on the average energy loss along the jet pass (solid), and one
utilizes stochastic energy loss $\Delta E(z)$ (dashed).
With realistic
3D medium evolution 
we find a striking 40-50\% reduction in elliptic flow 
compared to the case of simplified 1D dynamics.
Data from PHENIX (boxes) are also 
shown\cite{PHENIX_pi0_v2_y2010} to guide the eye.}
\label{Fig:v2_tuned}
\end{center}
\end{figure}
\begin{figure}[ht]
\leavevmode
\begin{center}
\includegraphics[height=50mm]{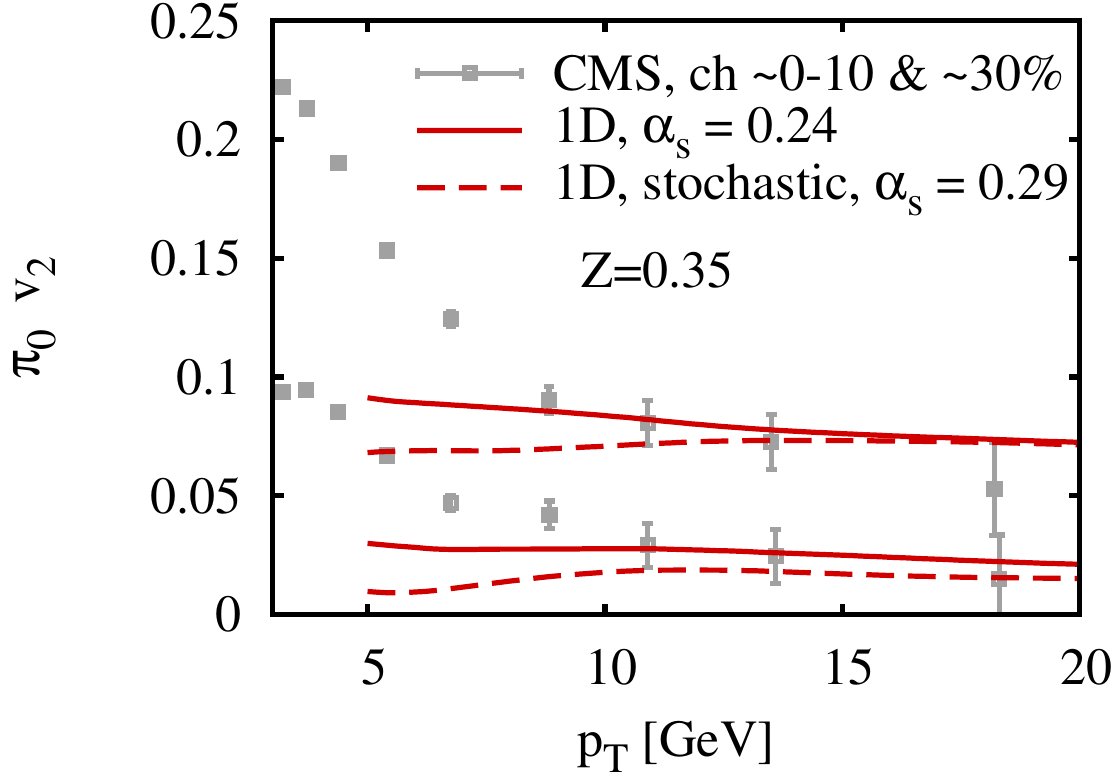}
\hskip 0cm
\includegraphics[height=50mm]{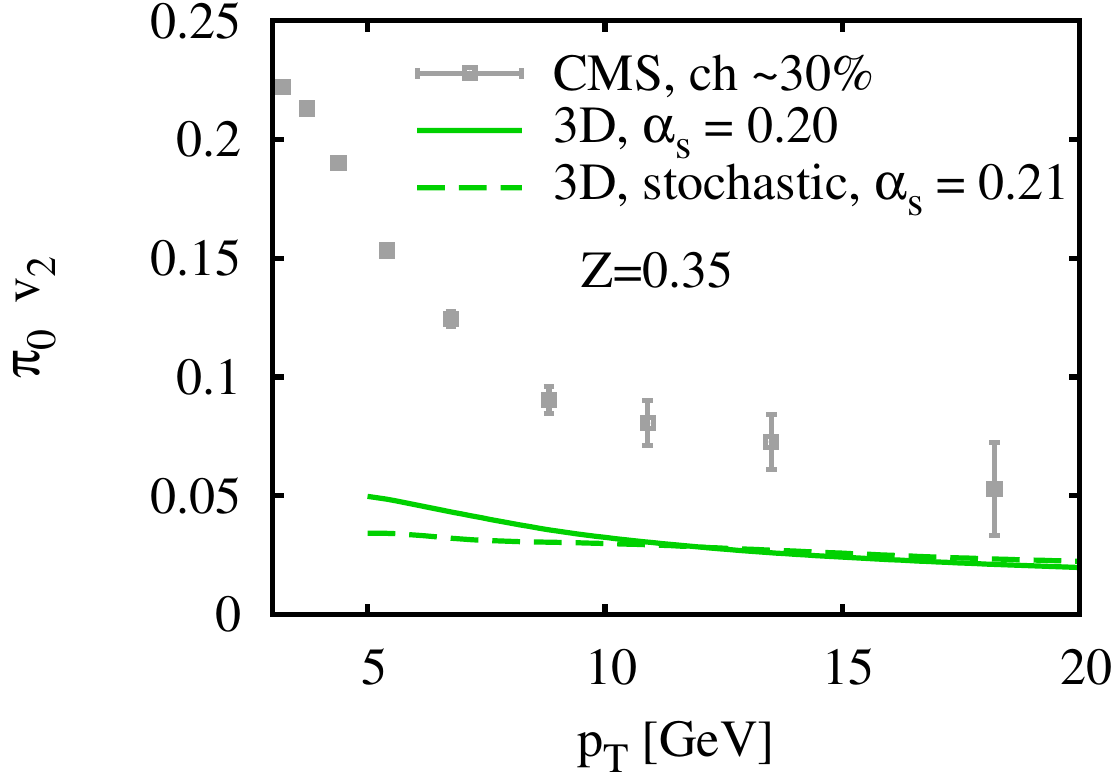}
\caption{Same as Fig.~\ref{Fig:v2_tuned} but for Pb+Pb at $\sqrt{s_{NN}} = 2.76$~TeV and with $\alpha_s$ tuned to reproduce $R_{AA} \approx 0.18$ 
at $p_T = 6$ GeV in central collisions. 
With realistic 3D medium evolution (right)
we only calculated $v_2$ for 
non-central collisions.
Transverse expansion
reduces $v_2$ dramatically at LHC energies as well.
Charged particle $v_2(p_T)$ data from CMS (boxes)
are also shown\cite{CMS_v2_ch_y2012} 
to guide the eye.}
\label{Fig:v2_LHC_tuned}
\end{center}
\end{figure}
We considered four scenarios
based on i) whether the medium is only undergoing Bjorken expansion 
(``1D'' as in \cite{CUJET1p0}) or transverse expansion as well (``3D'');
and ii) whether average energy loss is used or the stochastic $\Delta E(z)$.
Figure~\ref{Fig:R_AA} shows our results for $R_{AA}$ at RHIC, for the same
$\alpha_s = 0.29$. 
In the stochastic case energy loss effects are noticeably weaker, 
which is natural for convex parton spectra
(``curving up'' at high $p_T$). We also find that realistic transverse
expansion significantly enhances jet quenching, which is a generic GLV feature
coming from the
interference term in (\ref{GLV1}). Scatterings at large $z$
induce larger energy loss, and with a transversely expanding 
density profile there is higher chance to scatter further away from the production point than in the transversely static case.

Unfortunately,
without precise control over $\alpha_s$,
$R_{AA}$ alone cannot differentiate between these four scenarios.
As shown in Fig.~\ref{Fig:R_AA_tuned}, 
after a slight tuning of $\alpha_s$ to reproduce the
suppression in central collisions, differences in $R_{AA}$ 
largely disappear.
On the other hand, striking difference in $v_2$ remains between 1D and 3D
evolution at {\em both} 
RHIC and LHC energies, as shown in Figs.~\ref{Fig:v2_tuned}
and \ref{Fig:v2_LHC_tuned}. The strong $40-50$\% reduction of $v_2$ at 
high-$p_T$ with realistic transverse expansion 
is another generic consequence of interference in GLV. Scattering points
that lead to most energy loss are biased to occur away from the production
point and so later in time, by when the expansion makes the system
more cylindrical, reducing the spatial azimuthal asymmetry that 
drives elliptic flow.
We expect that this strong effect will be manifest in more full-fledged
GLV calculations as well, such as \cite{CUJET1p0}.

At the conference we also presented 
results based on scattering center ensembles from the transport
(not just density information)
but due to page limitations these will be written up elsewhere.

\section{Conclusions}

We investigated GLV energy loss using 
bulk medium evolution data from the covariant transport MPC. We find
that realistic transverse expansion strongly suppresses 
elliptic flow at high $p_T$ compared to calculations with transversely
``frozen'' profiles (as in \cite{CUJET1p0}). We argue that this is a generic
feature of GLV energy loss, raising the question whether GLV produces
too little elliptic flow at high $p_T$.

Transverse expansion also enhances the high-$p_T$ suppression, 
while fluctuations in energy loss with the location of scattering
centers reduce energy loss effects.
However, unlike for $v_2$, these effects 
nearly disappear once calculations are 
adjusted to reproduce $R_{AA}$ in central collisions.

\medskip
\noindent
{\bf Acknowledgements:} we thank A.~Buzzatti
and I.~Vitev for stimulating discussions.
This work was supported by the US DOE under grant
DE-PS02-09ER41665. Also, D.S. was partially supported by the JET Collaboration
(DOE grant DE-AC02-05CH11231).



%% References


%\bibliographystyle{elsarticle-num}



\end{document}